\documentclass{article}
\usepackage{amsfonts,amssymb, amsmath,mathrsfs}
\usepackage[english]{babel}

\newtheorem{proposition}{Proposition}

\def\g{{\gamma}}
\def\on#1#2{\mathop{\vbox{\ialign{##\crcr\noalign{\kern2pt}
$\scriptstyle{#2}$\crcr\noalign{\kern2pt\nointerlineskip}
\kern-2pt$\hfil\displaystyle{#1}\hfil$\crcr}}}\limits}

\newcommand{\dbyd}[2]{\dfrac{\partial #1}{\partial #2}}

\begin{document}

%%%%%%%%%%%% TITLE %%%%%%%%%%%%%%
\title{Integrable Euler top and nonholonomic Chaplygin ball}
\author{ A V Tsiganov \\
\it\small
St.Petersburg State University, St.Petersburg, Russia\\
\it\small e--mail:  andrey.tsiganov@gmail.com}

\date{}
\maketitle

\begin{abstract}
We  discuss the Poisson structures, Lax matrices, $r$-matrices, bi-hamiltonian structures, the variables of separation and other attributes of the modern theory of dynamical systems in application to  the integrable Euler top and to the nonholonomic Chaplygin ball.
\end{abstract}

\vskip0.1truecm

\section{Introduction}

The main aim of this paper is to prove that the integrable Euler top and the nonholonomic Chaplygin ball
are very similar dynamical systems like birds of a feather flock together. Thus, on example of these twins, we want to show how all the machinery developed for integrable systems can be carried to the theory of solvable nonholonomic systems.

The integrable Euler case of  rigid body motion with the fixed center of
mass (the Euler top) is relatively simple in the sense  that its equations of motion
do not linearize on Abelian surfaces, but on the elliptic curves. Of course, this does
not make  the Euler top entirely trivial \cite{fass}. A classical description of the Euler top can be found in any textbook on classical mechanics, see, for instance, \cite{am78,au96,bm05}.

The nonholonomic Chaplygin ball \cite{ch03} is that of a dynamically balanced 3-dimensional ball that rolls on a horizontal table without slipping or sliding. `Dynamically balanced' means that the geometric center coincides with the center of mass but
the mass distribution is not assumed to be homogeneous.  Because of the roughness of the table this ball cannot slip, but it can turn about the vertical axis without violating the constraints. There is a large body of literature dedicated to the Chaplygin ball, including the study of its generalizations. See \cite{bm01,bm08,du04,ekm04,ho09,jov09,koz83,koz85,mar86}. Of course, this list, as well as the bibliography of the present paper, is by far incomplete.

Section 2 starts by collecting some definitions and facts about Euler top and Chaplygin ball.  In Section 3 we will consider  Poisson structures associated with  the Chaplygin ball as deformations of the similar standard Poisson structures for the Euler top.
 Section 4 contains our main results about separability  at  zero level of the cyclic integral of motion.   In  bi-hamiltonian geometry  separability is an invariant geometric property of the distribution defined by mutually commuting independent integrals of motion.  In fact, there is  neither Hamilton-Jacobi equation, nor time which describes only   partial parametrization of geometric objects. We want to show how those standard bi-Hamiltonian geometric methods may be directly applied to the nonholonomic Chaplygin ball.

 In the second part of the paper we will discuss various deformations of the well-known integrable Hamiltonian systems which may be treated as generalizations of the Chaplygin ball.  In Section 5  we will briefly consider deformations of  integrable systems on cotangent bundles of the Riemannian manifolds and underline that  the main problem is the change of time which transfers a purely mathematical construction to the sensible physical model. For all these systems the deformations of initial Poisson bracket is trivial.  Finally in Section 6 we will give some examples of similar deformations on Lie algebra $e(3)$, when the  deformation of the initial Poisson bracket is nontrivial.

\section{Equations of motion}

Let $\g=(\g_1,\g_2,\g_3)$ and  $M=(M_1,M_2,M_3)$ be the two vectors of coordinates and momenta, respectively. We postulate that they satisfy to the following differential equations
\begin{equation}\label{m-eq}
\dot M=M\times \omega\,,\qquad \dot \g=\g\times \omega\,.
\end{equation}
For any  vector function $\omega$ on the dynamical variables $x=\g,M$ these equations in $\mathcal M=\mathbb R^3\times \mathbb R^3$ have the following integrals of motion
\begin{equation}\label{3int}
\mathcal H_1=(\g,\g),\qquad \mathcal H_2=(\g,M),\qquad \mathcal H_3=(M,M)\,.
\end{equation}
Six differential equations can be solved in quadratures if we know  four integrals and the Jacobi multiplier \cite{jac66}. So, we want to add some additional integral  to  the known integrals $\mathcal H_1,\mathcal H_2$ and $\mathcal H_3$ (\ref{3int}) and calculate the desired multiplier.

If we assume the existence of  the following additional integral of motion
\begin{equation}\label{f4-g}
\mathcal H_4=(M,\omega)
\end{equation}
one gets
\[\dfrac{d\mathcal H_4}{dt}=(M\times \omega,\omega)+(M,\dot{\omega})=(M,\dot{\omega})=0\,,\]
it means that the derivative $\dot{\omega}$ has to be perpendicular to $M$.
Below we stint ourselves by integrals (\ref{f4-g}) with
\begin{equation}\label{f4-2}
 \omega=\mathbf A_x M\,.
\end{equation}
In generic $\mathbf A_x$ is a matrix depending on variables $x=(\g,M)$, which has to satisfy to the equation
\[
(M,\dot\omega)=(\mathbf A_x^\top M,M\times \mathbf A_x M)+(M,\dot{\mathbf A}_xM)=0\,.
\]
This equation  can be replaced by the particular system of very simple equations
\begin{equation}\label{eq-C}
\mathbf A_x^\top=\mathbf{A}_x,\qquad\mbox{and}\qquad (M,\dot{\mathbf A}_x M)=0\,,
\end{equation}
which has constant  solution associated with the Euler top
\begin{equation}\label{eul-sol}
  \mathbf A_x=\mathbf A\,,\qquad \mathbf A=\left(
            \begin{array}{ccc}
              a_1 & 0 & 0 \\
              0 & a_2 & 0 \\
              0 & 0 & a_3
            \end{array}
          \right),\qquad a_k\in\mathbf R\,,
\end{equation}
 and solution associated  with the Chaplygin ball
\begin{eqnarray}\label{ch-sol}
  \mathbf A_x&=&\mathbf A_d\,,\qquad\mbox{where}\qquad  \mathbf A_d=\mathbf A +d\mathrm g(\g)\,\mathbf A\,\g \otimes\g\,\mathbf A\,,\\
\nonumber\\
\label{dot-a}
\dot{\mathbf A}_d&=&\mathrm g(\g)^2\mathbf A\left(\g\otimes\beta+\beta\otimes \g\right)\mathbf A,\qquad \beta=\bigl(\g-d(\g,\g)\mathbf A\g\bigr)\times \mathbf AM\,.
\end{eqnarray}
Here $\mathbf A$ is given by (\ref{eul-sol}) and  function $\mathrm g(\g)$ is equal to
\begin{equation}\label{g-fun}
\mathrm g(\g)=\dfrac{1}{1-d(\g,\mathbf A\g)}\,,
\end{equation}
so that $\mathbf A_d$ goes to $\mathbf A$ at $d\to 0$.

In the Euler-Poisson case, $M$ is the vector of the kinetic momentum,  $\omega$ is the angular velocity and $\g$  is the unit Poisson vector  \cite{au96,bm05}.  All these vectors are expressed in the so-called body frame,  its axes coincide with the principal inertia axes so that the corresponding tensor of inertia reads as
\begin{equation}\label{a-eul}
\mathbf J=\mathbf A^{-1}=\left(
            \begin{array}{ccc}
              J_1 & 0 & 0 \\
              0 & J_2 & 0 \\
              0 & 0 & J_3
            \end{array}
          \right)\,,\qquad J_k\leq J_i+J_j\,.
\end{equation}

In the  Chaplygin case  $M$ is the vector of angular momentum  of the ball with respect to the contact point,  $\omega$ is the angular velocity vector  of the rolling  ball and $\g$ is the unit normal vector to the plane at the contact point. As above all these vectors are expressed in the  body frame firmly attached to the ball.  If the  mass,  the inertia tensor and radius  of the rolling ball are denoted by  $m$, $\mathbf J$ and $a$, then matrix $\mathbf A$ in (\ref{ch-sol}) is equal to
\begin{equation}\label{a-ch}
\mathbf A=\left(
            \begin{array}{ccc}
              \dfrac{1}{J_1+d} & 0 & 0 \\
              0 & \dfrac1{J_2+d} & 0 \\
              0 & 0 & \dfrac{1}{J_3+d}
            \end{array}
          \right)\,,\qquad  d=ma^2\,.
\end{equation}
At $d\to 0$ equations of motion (\ref{m-eq}) and integrals of motion $\mathcal H_k$ for the Chaplygin ball coincide with  equations and integrals for the Euler top.  Of course,  it is purely mathematical and non-physical limit because  $J_i\to 0$ as $d\to 0$. So, from mathematical point of view, we can say that Chaplygin ball is a deformation of the Euler top with respect to parameter $d$.

In the Euler case matrix $\mathbf A$ and integral of motion  $\mathcal H_4$ are different from the
 matrix $\mathbf A$ and integral of motion  $\mathcal H_4$ in the Chaplygin case.  Nevertheless, for the brevity,  we will use common notations $\mathbf A$ and $\mathcal H_4$ in both cases where  it will not cause any  confusion.

Thus, we obtain the fourth integral of motion $\mathcal H_4$ (\ref{f4-2}) for the six equations (\ref{m-eq}) at $\omega=\mathbf A M$ and $\omega=\mathbf A_d M$. We proceed by showing that these dynamical systems are solvable in quadratures in framework of the Euler-Jacobi  last multiplier theory \cite{jac66}.  By definition, the Jacobi multiplier $\mu(x)$ of  (\ref{m-eq}) is a function on dynamical variables $x=\g,M$, which has to satisfy to the equation
\[
\sum_{i=1}^6 \dfrac{\partial}{\partial x_i}\, \mu(x)\,\dot{x}_i=0,\quad\Rightarrow\quad
\dot{\mu}(x)+\mu(x)\,\sum_{j=1}^3\left(\dfrac{\partial}{\partial \g_j}\,(\g\times \omega)_j +\dfrac{\partial}{\partial M_j}\, (M\times \omega)_j  \right)=0\,.
\]
For the solution $\mathbf A$ (\ref{eul-sol}) this equation is trivial
\begin{equation}\label{jac-mul1}
\dot{\mu}(x)=0\,\qquad\mbox{and}\qquad \mu(x)=\mu\equiv c,\qquad c\in\mathbb R,
\end{equation}
but for $\mathbf A_d$ one gets
\begin{equation}\label{jac-mul3}
2\mathrm g(\g)\dot{\mu}(x)-\mu(x)\dot{\mathrm g}(\g)=0\,,\qquad\mbox{and}\qquad \mu(x)= \mu_d\equiv c\sqrt{\mathrm g(\g)\,}\,.
\end{equation}
According to \cite{jac66}  the Jacobi's multiplier is some nontrivial function in the case of constrained systems only. The integrability conditions of the nonholonomic systems formulated by Kozlov \cite{koz85} include the preservation of  measure related with the Jacobi multiplier.

There are many other solutions of the system (\ref{eq-C}),  see  review \cite{bm08}.  For instance,  solution
\[\mathbf A_f=f(\g)\,\g \otimes\g\,,\]
depending on arbitrary function $f(\g)$ is associated with multiplier $\mu(x)=1$.  Solutions associated with
nontrivial Jacobi multiplier are given by  linear in variables $\g$ matrices
\begin{equation}\label{a-bc}
\mathbf A_{abc}=\mathbf A+\mathbf B (\g\otimes c+c\otimes \g)\mathbf B^\top\,,
\end{equation}
which satisfy (\ref{eq-C}), if we impose various restrictions on the numerical entries of matrices $\mathbf {A,B}$ and vector $c$.

To sum up, we can easily get a lot of additional integrals $\mathcal H_4$ (\ref{f4-2}) and the corresponding Jacobi multipliers of the equations (\ref{m-eq})  and, therefore,  we can solve these differential equations in  quadratures without any notion of the Hamilton structure, integrability by Liouville, the Poisson structure, the Lax matrices, classical $r$-matrices  etc.

However, this additional and in some sense redundant information can be useful in various applications, such as the perturbation theory, the quantization theory and so on. Below we reconstruct this information starting with  only integrals.

\section{The Poisson brackets.}

In this section our aim is to calculate the Poisson brackets for the given models without any assumptions on underlying Hamiltonian or conformally Hamiltonian structures of the equations of motion  \cite{bm01,bm08}. We will calculate the desired Poisson brackets only assuming that the foliation $H_i=\alpha_i$ is a direct sum of  symplectic and lagrangian foliations.

Let us consider the  manifold $\mathcal M$, dim $\mathcal M=n$, with coordinates $x=(x_1,\ldots,x_n)$.
The Jacobi last multiplier theorem \cite{jac66} ensures that $n$  equations
\begin{equation}\label{jac-eq}
\dfrac{dx_i}{dt}=X_i(x_1,\ldots,x_n)\,,\qquad i=1,\ldots,n.
\end{equation}
 are solvable in  quadratures if we have $n-2$ functionally independent integrals of motion $\mathcal H_k$ and the Jacobi multiplier $\mu$.

Let us suppose that $\mathcal M$  be a Poisson manifold endowed with a Poisson bivector $P$, so  that
the corresponding Poisson bracket reads as
\[
\{f,g\}=(Pdf,dg)=\sum_{i,j=1}^{dim \mathcal M}P_{ij}\dfrac{\partial f}{\partial x_i}\dfrac{\partial g}{\partial x_j}\,.
\]
In our examples,   all the symplectic foliation associated with the Poisson
bivector $P$ is rather regular.  Moreover, all the leaves are affine hyperplanes of codimension $k$, which  are the level sets of $k$ globally defined independent Casimir functions $C_j$
\[PdC_j=0,\qquad j=1,\ldots,k,\]
If  the symplectic foliation associated with $P$ is rather regular and $n=2m+k$, then we can determine the set of Hamiltonian systems
on $\mathcal M$
\begin{equation}\label{liuv-eq}
\dfrac{dx_i}{dt_j}=\{\mathcal H_j,x_i\}\,, \qquad i=1,\ldots,n,\qquad j=1,\ldots,n-2.
\end{equation}
The invariant volume form for all these Hamiltonian flows may be formally expressed as a square of symplectic form
on symplectic leaves
\[
\nu=\Omega^m\,,\qquad\mbox{where}\qquad \Omega=P_c^{-1}\,,
\]
and $P_c$ is restriction of $P$ on symplectic leaves.

Furthermore if one of these Hamiltonian systems (\ref{liuv-eq})  is integrable by Liouville we can compare the two integrable systems (\ref{jac-eq}) and (\ref{liuv-eq})  using change of time
\[ t\to t_j\]
and try to extract useful  information about the former system from  the  known properties of the latter Hamiltonian system.

 According to the Liouville theorem,  for the given Hamiltonian $\mathcal H_j$ equations of motion (\ref{liuv-eq}) are integrable in  quadratures on the symplectic leaves, if we have  $m$  functionally independent integrals of motion $H_1\equiv\mathcal H_j,H_2\ldots,H_m$ in the involution
\[
\{H_i,H_j\}=0\,,\qquad i,j=1,\ldots,m.
\]
 We can identify all the integrals  of the equations  (\ref{jac-eq}) with all the integrals  of (\ref{liuv-eq})
 \[
(\mathcal H_1,\ldots \mathcal H_{n-2})\sim (H_1,\ldots,H_m; C_1,\ldots,C_k)
 \]
 only at $m=2$, because $n=2m+k$ and $n-2=m+k$.

Regularity of symplectic foliation is closely related to the existence of the Jacobi multiplier $\mu$.
 One of the  global invariants in Poisson geometry is a modular class. It is  an obstruction to the existence of  measure in $\mathcal M$ which is invariant under all  hamiltonian flows \cite{kos85,vais93,wein97}.  For the manifold $\mathcal M$ endowed with a Poisson bivector $P$, its modular class is an element of the first Poisson cohomology group. In Section 3 we discuss some  elements of the second Poisson cohomology group and the corresponding  Poisson bivectors $P'$  compatible with $P$, which allows us to get  variables of separation without any additional information.

In generic case we can identify only one integral $H_1=\mathcal H_k$ and consider not only two-dimensional systems, but other systems as well. In this case the  integrals $\mathcal H_j$ could be generators of some algebra of integrals with respect to the bracket $\{.,.\}$, see theory of  superintegrable or noncommutative integrable systems  \cite{bol03,fass}.

In our case integrable by Euler-Jacobi  equations of motion (\ref{m-eq}) with integrals $\mathcal H_1,\ldots,\mathcal H_4$ and multiplier $\mu$
are integrable by Liouville after  an appropriate change of time, if there is the Poisson bivector $P$ such as
\begin{equation}\label{eq-p}
\begin{array}{ll}
\bullet\;[\![P,P]\!]=0,\qquad &\mbox{the Jacobi identity,}\\
\\
\bullet\;Pd\mathcal H_i=Pd\mathcal H_j=0,\qquad &\mbox{only two Casimir functions}\\
\\
\bullet\;\{\mathcal H_l,\mathcal H_m\}=0,\qquad &\mbox{the involution of the integrals\,.}
\end{array}
\end{equation}
Here $[\![.,.]\!]$ is the  Schouten bracket, $\mathcal H_1,\ldots, \mathcal H_4$ are the four integrals (\ref{3int}-\ref{f4-g}), and $(i,j,l,m)$ is the arbitrary permutation of $(1,2,3,4)$.

The first equation in (\ref{eq-p}) guaranties that $P$ is a Poisson bivector. In the second equation  we define  two Casimir elements $\mathcal H_{i}$ and $\mathcal H_j$ of $P$ and assume that rank$P=4$. It is a necessary condition because by fixing its values one gets the four dimensional symplectic phase space of our dynamical system. The third equation provides that the two remaining integrals $\mathcal H_l$ and $\mathcal H_m$ are in involution with respect to the Poisson bracket associated with $P$.

If we consider Chaplygin ball as a deformation of the Euler top, it is natural to  fix for the both systems the same Casimir functions
\begin{equation}\label{caz}
i=1,\,j=2\,,\qquad\Rightarrow\qquad \mathcal H_1=C_1\,,\qquad \mathcal H_2=C_2
\end{equation}
and integrals of motion
\begin{equation}\label{int-h}
l=3,\,m=4\,,\qquad\Rightarrow\qquad\mathcal H_4=H_1,\qquad \mathcal H_3=H_2\,.
\end{equation}
Solutions $P$ associated with another choice of Casimir functions  may be obtained from this solution
by using classical $r$-matrix theory, see  Section 3.3.

\subsection{The   linear in momenta Poisson bivectors.}
The system (\ref{eq-p}) has infinitely many solutions and, therefore, we have to narrow the search space and try to get some particular solutions only. In this Section we assume that the entries of $P$ are the linear functions in momenta $M$.
\begin{proposition}
In the  hypotheses  mentioned above the system of equations  (\ref{eq-p}) has the following linear in momenta solutions:
\begin{eqnarray}
&\mbox{Euler case:}\qquad& P=\left(\begin{array}{cc}0&\mathbf \Gamma\\ \mathbf \Gamma&
\mathbf M\end{array}\right)\,,\label{p-13}\\
\nonumber\\
&\mbox{Chaplygin case:}\qquad&P_d=\dfrac{1}{\sqrt{\mathrm g(\g)}}\,\left(\begin{array}{cc}0&\mathbf \Gamma\\ \mathbf \Gamma &
\mathbf M\end{array}\right)-d\sqrt{\mathrm g(\g)}\,(M,\mathbf A\g)\left(\begin{array}{cc}0&0\\ 0&
\mathbf \Gamma\end{array}\right)\,.
\nonumber
\end{eqnarray}
Here $\mathrm g(\g)$ is given by (\ref{g-fun}) and
\[
\mathbf \Gamma=\left(
                 \begin{array}{ccc}
                   0 & \g_3 & -\g_2 \\
                   -\g_3 & 0 & \g_1 \\
                   \g_2 & -\g_1 & 0
                 \end{array}
               \right)\,,\qquad
\mathbf M=\left(
                 \begin{array}{ccc}
                   0 & M_3 & -M_2 \\
                   -M_3 & 0 & M_1 \\
                   M_2 & -M_1 & 0
                 \end{array}
               \right)\,.
\]
\end{proposition}
The proof consists in the substitution of the linear in momenta anzats for entries of the Poisson bivector
 \[P_{ij}=\sum_k f_{ijk}(\g)M_k\]
into (\ref{eq-p}) and in the solution of the resulting algebro-differential equations with respect to unknown coefficients $f_{ijk}(\g)$.$\square$

The Poisson brackets between variables $x=\g,M$ look like
\begin{equation}\label{1-br}
\{M_i,M_j\}=\varepsilon_{ijk}M_k,\qquad \{M_i,\g_j\}=\varepsilon_{ijk}\g_k\qquad\{\g_i,\g_j\}=0,
\end{equation}
and
\begin{eqnarray}
\{M_i,M_j\}_d&=&\varepsilon_{ijk}\left(\dfrac{M_k}{\sqrt{\mathrm g(\g)}}-d\sqrt{\mathrm g(\g)}(M, \mathbf A\g)\g_k\right),\nonumber\\
\label{3-br}\\
 \{M_i,\g_j\}_d&=&\dfrac{\varepsilon_{ijk}\g_k}{\sqrt{\mathrm g(\g)}}\,,\qquad
\{\g_i,\g_j\}_d=0,\nonumber
\end{eqnarray}
Here $\varepsilon_{ijk}$ is a totally skew-symmetric tensor.

The first bracket $\{.,.\}$ is the well studied Lie-Poisson bracket on the Lie algebra $e^*(3)$,  whereas second bracket may be considered as  its deformation with respect to parameter  $d$, which preserves regular symplectic foliation.
The second Poisson bracket (\ref{3-br}) has been obtained   in \cite{bm01}.

Formally, in the both cases the Poisson bracket (\ref{1-br}-\ref{3-br})  allows us to rewrite the initial equations of motion (\ref{m-eq})  in the  Hamiltonian form
\begin{equation}\label{new-eq}
\dfrac{dx}{dt'}=\dfrac{dx}{\mu(x) dt}=\{H,x\},\qquad H=\dfrac{1}{2}\,\mathcal H_4\,,
\end{equation}
 after changing the  time variable including the corresponding  Jacobi multiplier
\begin{equation}\label{t-ch}
dt'=\mu(x)\, dt\,.
\end{equation}
where $\mu(x)=1$ or $\mu(x)=\mu_d$, respectively. In the Chaplygin case this transformation has been introduced in the Chaplygin work \cite{ch03} in order to get the  solutions as the functions of the time variable.

It is easy to see that for the Chaplygin ball  we can not directly identify  this transformation with canonical transformations of the extended phase space, which change time and the Hamilton function simultaneously  \cite{mak62,syng60,ts99,ts99a,ts01a}, see also discussion in \cite{bm08,ekm04,bl11,koz85}.

Of course, we can get similar Poisson bivectors for other solutions of (\ref{eq-C}) as well. For instance, if $\omega=\mathbf A_{ab}M$, where
\[
\mathbf A_{ab}=\left(
              \begin{array}{ccc}
                0 & 0 & 0 \\
                0 & 0 & 0 \\
                0 & 0 & a+b\g_3 \\
              \end{array}
            \right),
\]
is a matrix from the family (\ref{a-bc}), then  solution of (\ref{eq-p}) looks like
\begin{equation}\label{p-4}
P_{ab}=\dfrac{1}{\sqrt{ax_3+b}}\left[ \left(\begin{array}{cc}0&\mathbf \Gamma\\ \mathbf \Gamma &
\mathbf M\end{array}\right)-\dfrac{aM_3}{2(ax_3+b)}\left(\begin{array}{cc}0&0\\ 0&
\mathbf \Gamma\end{array}\right)              \right]\,.
\end{equation}
In this case the  equations of motion  (\ref{m-eq})  are integrable by   Euler-Jacobi theorem and
by the Liouville theorem after the corresponding change of  time.

\subsection{Properties of the linear Poisson bivectors.}
 The Poisson-Lichnerowicz cohomology of the Poisson manifold was defined in \cite{lih77}, and it provides a good framework to express the deformation and the quantization obstructions, see \cite{kos85,vais93,wein97}.

 Let us remind some necessary facts from the Poisson geometry.  The Poisson manifold $\mathcal M$ is a smooth (or complex manifold) endowed with the Poisson bivector $P$ fulfilling the Jacobi condition \[ [\![P,P]\!]=0\] with respect to the Schouten bracket on the algebra of the multivector fields on  $\mathcal M$. Other Poisson bivector $P'$ is compatible with  $P$ if any of its linear combination $P+\lambda P'$ is the Poisson bivector, i.e. if
\[ [\![P,P']\!]=0\,.\]
Bivectors  $P'$  are the 2-cocycles in the Poisson-Lichnerowicz cohomology defined by Poisson bivector $P$ on the Poisson manifold $\mathcal M$.  They must be compared with the bivectors
\begin{equation}\label{co-b}
P^{(X)}=\mathcal L_X\bigl(P\bigr)\qquad\Rightarrow\qquad [\![P,P^{(X)}]\!]=0
\end{equation}
which are the Lie derivative of $P$ along any vector field $X$ on $\mathcal M$.  Bivectors $P^{(X)}$
are 2-coboundaries  and 2-cocycles simultaneously. However not all cocycles are coboundaries.
If $X$ is such vector field that the Jacobi condition
\[[\![P^{(X)},P^{(X)}]\!]=0\] is satisfied, then  $P^{(X)}$ (\ref{co-b}) is called  the \textit{trivial} deformation of the Poisson bivector $P$.

Now let us go back to our physical models.  The first bivector $P$ (\ref{p-13}) is the well studied Lie-Poisson bivector on the Lie algebra $e^*(3)$ of Lie group $E(3)$ of Euclidean motions of $\mathbb R^3$. The second bivector  $P_d$ (\ref{p-13}) can be treated as its ``nonholonomic" deformation related with the Chaplygin ball.
\begin{proposition}
Bivector $P_d$ (\ref{p-13})  is  2-cocycle in the Poisson-Lichnerowicz cohomology defined by canonical Poisson bivector $P$ on  $e^*(3)$.
\end{proposition}

The proof is a straightforward calculation of the  Schouten bracket
\[
\qquad [\![P,P_d]\!]=0\,.
\]
Recall that the Schouten bracket $[\![R,Q]\!]$ of two bivectors $R$ and $Q$
is trivector and its entries in local coordinates $x$ look like
\begin{equation}\label{sh-br}
[\![R,Q]\!]^{ijk}=-\sum\limits_{m=1}^{dim\,\mathcal  M}\left(Q^{mk}\dfrac{\partial R^{ij}}{\partial x^m}
+R^{mk} \dfrac{\partial Q^{ij}}{\partial x^m}+\mathrm{cycle}(i,j,k)\right).
\end{equation}
 $\square$

\begin{proposition}
In the generic case the Poisson bivector
\begin{equation}\label{p3}
 P_d=\dfrac{1}{\sqrt{\mathrm g(\g)}}\,\left(\begin{array}{cc}0&\mathbf \Gamma\\ \mathbf \Gamma &
\mathbf M\end{array}\right)-d\sqrt{\mathrm g(\g)}\,(M,\mathbf A\g)\left(\begin{array}{cc}0&0\\ 0&
\mathbf \Gamma\end{array}\right)
\end{equation}
is a nontrivial deformation of the standard Lie-Poisson bivector  $P$, which is a sum of two Lie derivatives of $P$
\begin{equation}\label{p3-l}
P_d=\mathcal L_Y\bigl(P\bigr)+\dfrac{d\sqrt{\mathrm g(\g)}\,(\g,M)}{2}\,\mathcal L_Z\bigl(P\bigr).
\end{equation}
Here entries  of the vector fields $Y=\sum Y^j\partial_j$ and $Z=\sum Z^j\partial_j$ are given by
\begin{equation}\label{yz-field}
Y^{i}=Z^{i}=0,\qquad
Y^{i+3}=-\dfrac{M_j}{\sqrt{\mathrm g(\g)}},\quad
Z^{i+3}=\Bigl(\bigl(\mathrm{tr}\mathbf A\cdot\mathbf{Id}-\mathbf A\bigr)\,\g\Bigr)_i,
\qquad i=1,2,3\,.
\end{equation}
\end{proposition}

In order to prove first part of this proposition we have to try to solve the following  equation
\begin{equation}\label{eq-X}
P_d=\mathcal L_X(P)\,,\end{equation}
with respect to unknown $X$.  Recall that in local coordinates the Lie derivative of a bivector $P$
along a vector field $X$ reads
\[
\Bigl(\mathcal L_{X}(P)\Bigr)^{ij}=\sum\limits_{k=1}^{dim M}\left(X^k\dfrac{\partial  P^{ij}}{\partial  x^k}
-P^{kj}\dfrac{\partial  X^i}{\partial  x^k} -P^{ik}\dfrac{\partial X^j}{\partial  x^k}\right).
\]
Using the modern software for symbolic calculations it is easy to prove that
entries of (\ref{eq-X}) form the inconsistent system of differential equations. It means that (\ref{eq-X}) is infeasible equation and, therefore, cocycle $P_d$ is no coboundary. The second part of  the proposition is verified by direct calculations.
 $\square$

It is easy to see, that if $(\g,M)=0$ then $P_d=\mathcal L_Y\bigl(P\bigr)$ is a \textit{trivial} deformation with all the pleasant mathematical and physical  consequences, see \cite{lih77,vais93} and \cite{bm08,ch03} respectively.

In the finite-dimensional case local Poisson geometry begins with the
splitting theorem, which says that in the neighborhood of any point in the Poisson mani\-fold $\mathcal M$, there are coordinates
$(q_1,\ldots, q_m$, $p_1,\ldots,p_m$, $C_1,\ldots,C_k )$  such as
\[
P	=\sum_{i=1}^m\dbyd{}{q_i}\wedge\dbyd{}{p_i}+\dfrac12
\sum_{i,j=1}^{k}\varphi_{ij}(C)\dbyd{}{C_i}\wedge\dbyd{}{C_j}
\ \ \ \mbox{ and } \ \ \ \varphi_{ij}(0)=0\ .
\]
So,  if the compatible bivectors $P$ and $P'$ have a common set of Casimirs $C_1,\ldots, C_k$, we can
identify the Darboux coordinates $(q,p)$ of $P$ with the Darboux coordinates $(q',p')$ of $P'$ and obtain the
local map $\phi: \mathcal M\to \mathcal M$, which pulls back $P'$  to $P$.
In Section 4.3 we prove that in our case this local map $P_d\to P$ can be extended to the global one at $C_2=0$.

If we come back to the general theory, the second Poisson-Lichnerowicz cohomology group $\mathcal H^2_{P}$ on $\mathcal M$ is precisely the set of bivectors $P'$ solving $[\![P,P']\!]=0$ modulo the solutions of the form $P^{(X)}=\mathcal L_X(P)$.
We can interpret  $\mathcal H^2_{P}$ as the space of infinitesimal deformations of the Poisson structure modulo trivial deformations. We should keep in mind that  cohomology reflects the topology of the leaf space and the variation in the symplectic structure as one passes from one leaf to another  \cite{lih77,kos85,vais93,wein97}.

\subsection{The $r$-matrices}
It is known that equations (\ref{m-eq}) can be  rewritten in the Lax form
\begin{equation}\label{lax-ch}
\dfrac{d\mathbf L}{dt}=[\mathbf L,\mathbf \Omega],\qquad \mathbf L=\mathbf M+\dfrac{\mathbf \Gamma}{\lambda},\qquad\qquad \lambda\in\mathbb R\,.
\end{equation}
if we identify
$(\,\mathbb R^3,\times)$ and  $(so(3),[.,.])$ by using a well known isomorphism
\begin{equation}\label{trans-M}
 z=\left(z_1,z_2,z_3\right)\to \mathbf Z=\left(\begin{matrix}
            0 & z_3 & -z_2 \\
            -z_3 & 0 & z_1 \\
            z_2 & -z_1 & 0
          \end{matrix}\right),
\end{equation}
where $\times$ is a cross product in $\mathbb R^3$ and  $[.,.]$ is a matrix commutator in $so(3)$.
Another possibility is to use $4\times 4$ antisymmetric matrices
\[
\mathbf M=\left(
            \begin{array}{cccc}
              0 &M_3 &-M_2 & 0 \\
              -M_3 & 0 & M_1 & 0 \\
              M_2 & -M_1 & 0 & 0 \\
              0 & 0 & 0 & 0 \\
            \end{array}
          \right)\,,\qquad \mathbf \Omega=\left(
            \begin{array}{cccc}
              0 &\omega_3 &-\omega_2 & 0 \\
              -\omega_3 & 0 & \omega_1 & 0 \\
              \omega_2 & -\omega_1 & 0 & 0 \\
              0 & 0 & 0 & 0 \\
            \end{array}
          \right)
\]
and symmetric matrix
\[
\mathbf \Gamma=\left(
            \begin{array}{cccc}
              0 &0 &0 & \g_1 \\
              0 & 0 & 0 & \g_2 \\
              0 & 0 & 0 & \g_3 \\
              \g_1 & \g_2 & \g_3 & 0 \\
            \end{array}
          \right)\,.
\]
In the both cases bilinear Lax matrix $\mathbf L(\lambda)$ (\ref{lax-ch}) belongs to a huge family of Lax matrices described in the book \cite{rs} , see also \cite{sn02}. Below we will consider only $3\times 3$ Lax matrices.

The Lax equation  implies that the spectral invariants of the Lax matrix $\mathbf L(\lambda)$ are conserved
quantities under the Hamiltonian evolution, but their involutivity and functionally independence  must be checked case by case.
In a noteworthy paper \cite{bv90}, Babelon and Viallet showed
that if all the spectral invariants  of the $m\times m$  matrix $\mathbf L(\lambda)$ are in involution  with respect to some Poisson bracket $\{.,.\}$ on a given phase space,
then there is  a matrix $r_{12}(\lambda,\mu)$ of order $m^2\times m^2$ such that the
Poisson brackets between the entries of $L$ are represented in the commutator form
\begin{equation}
\{\,\on{\mathbf L}{1}(\lambda),\,\on{\mathbf L}{2}(\mu)\}_k= [r_{12}(\lambda,\mu)\,,\,\on{\mathbf L}{1}]-[r_{21}(\lambda,\mu)\,,\,\on{\mathbf L}{2}(\mu)\,]\,. \label{rrpoi}
\end{equation}
Here $\on{\mathbf L}{1}(\lambda)=\mathbf L(\lambda)\otimes \mathbf{Id}\,,~\on{\mathbf L}{2}(\mu)=\mathbf{Id}\otimes \mathbf L(\mu)$ and
$r_{12}(\lambda,\mu)$ is a classical $r$-matrix and
\[
r_{21}(\lambda,\mu)=\mathbf Pr_{12}(\mu,\lambda)\,,
\]
where $\mathbf P$ is a permutation operator: ${\Pi}x\otimes y =y\otimes x$,\ $\forall x,y\in \mathbb C^m$ \cite{rs}.

In our case for the Euler top and for the Chaplygin ball we have one Lax matrix $\mathbf L(\lambda)$ and two different Poisson brackets.  First bracket $\{.\,,.\}$ (\ref{1-br}) associated with the  bivector $P$ (\ref{p-13}) yields the standard $r$-matrix
\begin{equation}\label{r-eul}
r_{12}(\lambda,\mu)=\dfrac{\mu}{\mu-\lambda}\sum_{i=1}^3 \mathbf S_i\otimes \mathbf S_i\,.
\end{equation}
 Here $\mathbf S_i$ form a basis in the space of $3\times 3$ antisymmetric matrices
\[
\mathbf S_1=\left(\begin{matrix}
            0 & 0 & 0 \\
            0 & 0 & 1 \\
            0 & -1 & 0
          \end{matrix}\right),\qquad
\mathbf S_2=\left(\begin{matrix}
            0 & 0 & -1 \\
            0 & 0 & 0 \\
            1 & 0 & 0
          \end{matrix}\right),\qquad
\mathbf S_3=\left(\begin{matrix}
            0 & 1 & 0 \\
            -1 & 0 & 0 \\
            0 & 0 & 0
          \end{matrix}\right)\,.
\]
For the nonholonomic bracket $\{.,.\}_d$ (\ref{3-br}) associated with the bivector $P_3$ (\ref{p-13}) the $r$-matrix will be a more complicated dynamical $r$-matrix.

\begin{proposition}
The Lax matrix for nonholonomic Chaplygin ball (\ref{lax-ch}) satisfies the linear $r$-matrix algebra  (\ref{rrpoi}) with the following $r$-matrix
\begin{equation}\label{r-ch}
r_{12}(\lambda,\mu)=\dfrac{\mu}{\mu-\lambda}\left(\dfrac{1}{\sqrt{\mathrm g(\g)}}-d\lambda\sqrt{\mathrm g(\g)}\,(M,\mathbf A\g)\right)\sum_{i=1}^3 \mathbf S_i\otimes \mathbf S_i\,.
\end{equation}
\end{proposition}

The proof is straightforward verification of (\ref{rrpoi}) for the given Lax matrix..
 $\square$
Usually, the bilinear Lax matrices  (\ref{lax-ch})  are not  very useful in integration of equations of motion.  However,  they are very effective in various geometric applications, for instance see one of the latest applications in discrete differential geometry \cite{kp11}.

In bi-Hamiltonian geometry we can use bilinear Lax matrices in order to get solutions of (\ref{eq-p}) associated with another choice of the Casimir elements. Namely,    if  $r_{12}$ is classical $r$-matrix and $\varphi$ is intertwining  operator, then $r_{12}\circ \varphi$  is also a classical $r$-matrix  \cite{rs}. For a given matrix $r_{12}$ the $r$-matrices $r_{12}\circ \varphi$ form a linear  Lie pencil, which generates a family of compatible Lie-Poisson brackets.  For instance,  if we take  trivial  intertwining  operators from \cite{rs}
\[
\varphi_1=\mu^{-1}\,,\qquad\mbox{and}\qquad \varphi_2=\mu^{-2}
\]
and $r$-matrix (\ref{r-eul}), then one gets  the following well-known Poisson bivectors  for the Euler top
\[
P_1=\left(\begin{array}{cc} \mathbf \Gamma&0 \\ 0 &0 \end{array}\right)\,,\qquad\mbox{and}\qquad
P_2=\left(\begin{array}{cc} \mathbf M&0 \\ 0 &0 \end{array}\right)\,,
\]
see \cite{mor96}. Construction of other  brackets on $e^*(3)$ associated with classical $r$-matrices (\ref{r-eul}-\ref{r-ch}) and more sophisticated relations between various classical $r$-matrices \cite{ts07a} will be discussed in forthcoming publication.

\section{Separation of variables at $(\g,M)=0$.}

Now we address the problem of  separation of variables  within the theoretical scheme of bi-hamiltonian geometry \cite{fp02,ts09a,ts09b}.  According to \cite{ch03}  we can start with the case $(\g,M)=0$ and then reduce  the generic case to this particular one.

In geometry, instead of  an additive separation of variables in the partial differential equation called the Hamilton-Jacobi equation,  we have some invariant  geometric property of the Lagrangian distribution defined by  $m$ independent functions  $H_1,\ldots,H_m$ \cite{fp02,ts09a,ts09b}.

Namely, an $m$-tuple  $H_1,\ldots,H_m$ of functionally independent functions  defines a separable foliation on symplectic leaves of  $\mathcal M$, if there are   variables of separation $(  q_1,\dots,  q_m,  p_1,\dots,  p_m)$ and $m$ separated  relations of the form
\begin{equation}
\label{seprelint}
\Phi_i=0\ ,
\quad\mbox{with}\quad\det\left[\frac{\partial \Phi_i}{\partial H_j}\right]
\not=0\,.
\end{equation}
It simple means that the common level surfaces of $H_1,\ldots,H_m$ form rather regular foliation of symplectic leaves and every leaf of this lagrangian foliation may be represented as a direct product of one-dimensional geometric objects defined by separated relations  (\ref{seprelint}).

Now let us remind how to get variables of separation in framework of the bi-Hamiltonian geometry. The bi-Hamiltonian manifold $\mathcal M$ is a smooth  (or complex) manifold endowed with two compatible Poisson bivectors $P$ and $P'$. Dynamical systems on $\mathcal M$
with the integrals of motion in  involution with respect to the both brackets
\begin{equation}\label{inv-H}
\{H_i,H_j\}=\{H_i,H_j\}'=0,\qquad  i,j=1,\ldots,m,
\end{equation}
are called  bi-integrable systems \cite{ts09a,ts09b}. The bi-involutivity of the integrals of motion (\ref{inv-H}) is equivalent to the existence of  the control matrix $ F$ defined by
\begin{equation}\label{f-mat}
P'dH_i=P\sum_{j=1}^m   F_{ij}\,dH_j,\qquad i=1,\ldots,m.
\end{equation}
The eigenvalues $(q_1,\ldots,q_m)$ of  $F$ are the coordinates of separation, whereas
the suitable normalized left eigenvectors of $F$ form the generalized St\"ackel matrix $S$
\[
 F=S^{-1}\,\mbox{diag}\,(q_1,\ldots,q_m)\, S
 \]
which defines the separation relations
\begin{equation}
\label{stseprel}
\Phi_i=\sum_{j=1}^n S_{ij}(q_i,p_i) H_j+U_i(q_i,p_i)=0\ ,\qquad
i=1,\dots,m\ .
\end{equation}
Here the entries of  St\"ackel matrix $S_{ij}$ and the St\"ackel potentials
$U_i$ depend only on one pair  $(q_i,p_i)$ of the canonical variables of separation, Casimir functions $C_j$ and, in generic case, on the integrals of motion \cite{fp02,ts09a,ts09b}.

In our case the St\"ackel matrix  and the St\"ackel potentials
depend only on variables of separation, and it allows us to calculate the canonical transformation from the initial variables $\g,M$ to the variables of separation explicitly.

\subsection{Darboux-Nijenhuis coordinates}
In order to get variables of separation according to the general usage of bi-hamiltonian geometry  firstly  we have to calculate the bi-hamiltonian structure for the given  systems with integrals of motion $H_1,H_2$ (\ref{int-h}) on manifold $\mathcal M$ with the canonical Poisson bivector $P$ (\ref{p-13}) and its deformation $P_d$.

\begin{proposition}
Let us introduce two vector fields $X=\sum X^j\partial_j$ and $X_d=\sum X_d^j\partial_j$, with the following entries:
\begin{equation}\label{field-ell}
\begin{array}{lll}
X^i=0,\qquad & X^{i+3}=\Bigl[\g\times\mathbf A (\g\times M)\Bigr]_i,\quad & i=1,2,3\,.\\ \\
X_d^i=0,\qquad & X_d^{i+3}=\Bigl[\g\times\mathbf A_d (\g\times M)\Bigr]_i\,.&
\end{array}
\end{equation}
The Poisson bivectors
\begin{equation}\label{p-p}
P'=\mathcal L_{X}P\qquad\mbox{and}\qquad
P'_d=\mathcal L_{X_d}P_d
\end{equation}
are compatible with the bivectors $P$ and $P_d$ (\ref{p-13}) respectively.  Bivectors  (\ref{p-p})  have common symplectic leaves
\begin{equation}\label{com-caz}
P'dC_1=P'_d dC_1=0,\qquad P'dC_2=P'_d dC_2=0,
\end{equation}
 whereas integrals of motion $H_{1,2}$ (\ref{int-h}) are in the bi-involution
\begin{equation}\label{inv-F}
\{H_1,H_2\}=\{H_1,H_2\}'=0\,,\qquad \{H_1,H_2\}_d=\{H_1,H_2\}'_d=0
\end{equation}
with respect to the corresponding Poisson brackets at  $(\g,M)=0$ only .
\end{proposition}

The proof   is a straightforward verification of the corresponding Schouten and Poisson brackets in local coordinates.
 $\square$
Thus, we proved that the Euler top and the nonholonomic Chaplygin ball are bi-integrable systems at $(\g,M)=0$.
At second step we have to calculate  the corresponding   control matrices $F$ and $F_d$ defined by (\ref{f-mat}).
For the Euler top we have
\begin{equation}\label{fk-mat}
F=\left(
             \begin{array}{cc}
               0 & (\mathbf A^\vee \g,\g) \\
               -(\g,\g) & \Bigl(\bigl(\mathrm{tr}\mathbf A\cdot\mathbf{Id}-\mathbf A\bigr)\g,\g\Bigr) \\
             \end{array}
           \right)
\end{equation}
and similar to  the Chaplygin ball
\begin{equation}\label{fk-mat2}
F_d=\left(
             \begin{array}{cc}
               0 & (\mathbf A_d^\vee \g,\g) \\
               -(\g,\g) & \Bigl(\bigl(\mathrm{tr}\mathbf A_d\cdot\mathbf{Id}-\mathbf A_d\bigr)\g,\g\Bigr) \\
             \end{array}
           \right)\,.
\end{equation}
Here  $\mathbf A^\vee=({\det \mathbf A})\,{\mathbf A}^{-1}$ is  adjoint or cofactor matrix.

The Darboux-Nijenhuis coordinates associated with the bivectors (\ref{p-p}) and control matrices (\ref{fk-mat}-\ref{fk-mat2})   are the roots of their characteristic polynomials
\begin{equation}\label{sep-coord}\begin{array}{l}
\tau(\lambda)=\lambda^2-\Bigl(\bigl(\mathrm{tr}\mathbf A\cdot\mathbf{Id}-\mathbf A\bigr)\g,\g\Bigr)\lambda+(\g,\g)
({\mathbf A}^\vee \g,\g)=0\,,\\
\\
\tau_d(\lambda)=\lambda^2-\Bigl(\bigl(\mathrm{tr}\mathbf A_d\cdot\mathbf{Id}-\mathbf A_d\bigr)\g,\g\Bigr)\lambda+(\g,\g)
({\mathbf A}_d^\vee \g,\g)=0\,.
\end{array}
\end{equation}
By definition \cite{fp02},  the Darboux-Nijenhuis variables are canonical with respect to the symplectic form $\Omega$ associated with first bivector $P$ and put the recursion operator $N=P'P^{-1}$ in diagonal form on symplectic leaves of $\mathcal M$.

Below we prove that these Darboux-Nijenhuis coordinates (\ref{sep-coord}) are the variables of separation for the bi-lagrangian foliation defined by integrals $H_{1,2}$ (\ref{int-h}) on symplectic leaves of $e^*(3)$ fixed by $C_1=1$ and $C_2=0$.

It is easy to see that at $(\g,M)=0$ the passage from the Euler top to the nonholonomic Chaplygin ball consists
of the replacement of the constant matrix $\mathbf A$ (\ref{eul-sol}) on dynamical one $\mathbf A_d$ (\ref{ch-sol}) in the equations of motion (\ref{m-eq}), in the Hamiltonian $H_1=(M,\mathbf AM)$ and equations (\ref{field-ell},\ref{fk-mat},\ref{sep-coord}) only.  Similar to geometric quantization theory,  simplicity of this deformation is a sequence  of the equation (\ref{p3-l})
\[
P_d=\mathcal L_Y(P),
\]
 properties of the Lie derivative $\mathcal L$ and of the vector fields $Y$ (\ref{yz-field}) and $X_d$ (\ref{field-ell}).

\subsection{Elliptic coordinates}
At $C_2=0$ and $C_1=1$ we can  identify the corresponding symplectic leaf of $\mathcal M=e^*(3)$ with the cotangent bundle $T^*\mathbb S$ of the unit two-dimensional Poisson sphere \cite{bm05}.

 If we put $C_1=(\g,\g)=1$, then,  dividing characteristic polynomials (\ref{sep-coord}) on det$(\mathbf A-\lambda\,\mathbf{Id})$ we get the standard definitions of elliptic coordinates $u,v$ on the sphere and their nonholonomic  deformations $\mathrm u,\mathrm v$:
\begin{equation}\label{ell-q}
e(\lambda)=\dfrac{\g_1^2}{\lambda-a_1}+\dfrac{\g_2^2}{\lambda-a_2}+\dfrac{\g_3^2}{\lambda-a_3}
=\dfrac{(\lambda-u)(\lambda-v)}{(\lambda-a_1)(\lambda-a_2)(\lambda-a_3)},\qquad \end{equation}
and
\begin{eqnarray}\label{nhell-q}
e_d(\lambda)&=&\mathrm g(\g)\,\left(\dfrac{\g_1^2(1-da_1)}{\lambda-a_1}+\dfrac{\g_2^2(1-da_2)}{\lambda-a_2}+
\dfrac{\g_3^2(1-da_3)}{\lambda-a_3}\right)\\
\nonumber\\
&=&\dfrac{(\lambda-\mathrm{u})(\lambda-\mathrm{v})}{ (\lambda-a_1)(\lambda-a_2)(\lambda-a_3)}\,,\nonumber
\end{eqnarray}
respectively.  Function $\mathrm g(\g)$ in coordinates $\mathrm u,\mathrm v$ reads as
\begin{equation}\label{gg-uv}
\mathrm g(\g)=\dfrac{(1-d\mathrm{u})(1-d\mathrm{v})}{(1-da_1)(1-da_2)(1-da_3)}\,.
\end{equation}
Equation (\ref{ell-q}) is a standard definition of the elliptic coordinates $u,v$ on the unit sphere, whereas equation (\ref{nhell-q}) determines the nonholonomic elliptic coordinates $\mathrm{u},\mathrm{v}$. In \cite{bm08b}, the coordinates of the form (\ref{nhell-q}) are used for the integration of the sphere-sphere problem under the name quasi-spheroconical coordinates, see discussion in \cite{ts11a}.

We have to point out that our aim is the {calculation} of the  variables of separation without any additional assumptions. Thus, we have to calculate the conjugated momenta $p_u,p_v$ and $\mathrm{p_u},\mathrm{p_v}$ in framework of bi-hamiltonian geometry. It is easy to prove that in our  case we have identical St\"ackel matrices
\[
S=\left(
            \begin{array}{cc}
              1 & 1 \\
              -u & -v
            \end{array}
          \right)\,\qquad\mbox{and}\qquad
 S_d=\left(
            \begin{array}{cc}
              1 & 1 \\
              -\mathrm{u} & -\mathrm{v}
            \end{array}
          \right)\,,
\]
and identical St\"ackel potentials
\begin{equation}\label{st-pot}
\begin{array}{l}
 U_1^{(1)}=u\,H_2-H_1\,, \quad
 U_2^{(1)}=v\,H_2-H_1\,,\\
\\
{U}_1^{(3)}=\mathrm{u}\,H_2-H_1\,, \quad
{U}_2^{(3)}=\mathrm{v}\,H_2-H_1\,,
\end{array}
\end{equation}
where
\[
\qquad H_1=\mathcal H_4=(M,\omega) ,\qquad H_2=\mathcal H_3=(M,M) .
\]
According to \cite{ts09a,ts09b}, the notion of the St\"ackel potentials allows us to find unknown conjugated momenta using the Poisson brackets only.

For instance, the following recurrence chain of the Poisson brackets
\begin{equation}\label{rrel2}
\phi_1=\{u,U_1^{(1)}\}_1,\qquad \phi_2=\{u,\phi_1\}_1,\ldots,\quad \phi_i=\{u,\phi_{i-1}\}_1
\end{equation}
breaks down on the third step $\phi_3=0$. It means that ${U}_1^{(1)}(u,p_u)$ is the second order polynomial in momentum $p_u$ and, therefore, we can define this unknown momentum in the following way
\begin{equation}\label{pkow-def1}
p_u=\dfrac{\phi_1}{\phi_2}=\sum_{ijm} \varepsilon_{ijm}\,\dfrac{\g_j\g_m(a_j-a_m)(a_i-u)}{2(u-v)(a_j-u)(a_m-u)}\,M_i
\end{equation}
up to the  canonical transformations $p_u\to p_u+ f(u)$. As above,  $\varepsilon_{ijk}$ is a totally skew-symmetric tensor.

Similar calculation with  $U^{(1)}_2(v,p_v)$ yields to the  definition of the second momentum $p_v$.
In  nonholonomic case we can perform completely identical calculations too.
The results obtained so far can be summarized in the following statement.
\begin{proposition}
The initial coordinates $x=\g,M$ are expressed via elliptic coordinates $u,v$ and $p_u,p_v$
\begin{eqnarray}
\g_i&=&\sqrt{\dfrac{(u-a_i)(v-a_i)}{(a_j-a_i)(a_m-a_i)}},\qquad i\neq j\neq m\,,\nonumber\\
\label{ftrans-ell}\\
M_i&=&\dfrac{2\varepsilon_{ijm}\g_j\g_m(a_j-a_m)}{u-v}\Bigl((a_i-u)p_u-(a_i-v)p_v\Bigr)\,,
\nonumber
\end{eqnarray}
In terms of the nonholonomic elliptic coordinates $\mathrm{u},\mathrm{v}$ and $\mathrm{p}_u,\mathrm{p}_v$   the same variables look like
{\setlength\arraycolsep{1pt}
\begin{eqnarray} \label{ftrans-nell}
\g_i&=&\sqrt{\dfrac{(1-da_j)(1-da_m)}{(1-d\mathrm{u})(1-d\mathrm{v})}}
\,\cdot\,\sqrt{\dfrac{(\mathrm{u}-a_i)(\mathrm{v}-a_i)}{(a_j-a_i)(a_m-a_i)}}\,
,\qquad i\neq j\neq m\,,\\\
\nonumber\\
M_i&=&\dfrac{2\varepsilon_{ijk}\g_j\g_k(a_j-a_k)\sqrt{\mathrm g(\g)}}{\mathrm{u}-\mathrm{v}}\,
\Bigl((a_i-\mathrm{u})(1-d\mathrm{u})\mathrm{p}_u-(a_i-\mathrm{v})(1-d\mathrm{v})\mathrm{p}_v\Bigr)\,.
\nonumber
\end{eqnarray}}
where $\mathrm g(\g)$ is given by (\ref{gg-uv}).
\end{proposition}
  It is simple combination of definitions of the Casimir functions $C_{1,2}$,  coordinates  (\ref{ell-q}-\ref{nhell-q}),   momentum (\ref{pkow-def1}) and   other momenta.  $\square$

\subsection{The reduction of the Poisson brackets}
The two sets of variables of separation  $u,v,p_u,p_v$ and  $\mathrm{u}, \mathrm{v},\mathrm{p}_u,\mathrm{p}_v$ are the Darboux variables with respect to the brackets $\{.,.\}$ (\ref{1-br}) and  $\{.,.\}_d$ (\ref{3-br}) on the symplectic leaf $C_1=1$ and $C_2=0$. Of course,  we can identify these variables and  get the diffeomorphism $\phi:\mathcal M \to \mathcal M$, which  pulls back the nonholonomic bracket $\{.,.\}_d$ to the standard Lie-Poisson bracket $\{.,.\}$ on the Lie algebra $e^*(3)$ at $C_2=0$.
\begin{proposition}
At $(\g,M)=0$ the Poisson bracket $\{.,.\}_d$ (\ref{3-br}) between the variables $\g,M$ coincides with the Lie-Poisson bracket $\{.,.\}$ (\ref{1-br})  between the variables
\begin{eqnarray}
\hat{\g}_j&=&\sqrt{\mathrm g(\g)\,\Bigl(1-d(\g,\g)\,a_j\Bigr)}\,\g_j\,,\qquad j=1,2,3,\nonumber\\
\label{ch-evar}\\
\hat{M}_j&=&\sqrt{\dfrac{1}{\prod_{i\neq j}\Bigl(1-d(\g,\g)\,a_i\Bigr)}}\left(
\dfrac{M_j}{\sqrt{\mathrm g(\g)}}+d\sqrt{\mathrm g(\g)}\,(M,\mathbf A\g)\,\g_j
\right)\,.\nonumber
\end{eqnarray}
This mapping identifies variables $\mathrm{u}, \mathrm{v}$ with the usual elliptic coordinates $u,v$ on the sphere, which were used by Chaplygin \cite{ch03}.
\end{proposition}
  It is a direct sequence of the previous Proposition. $\square$

So, at $(\g,M)=0$ we can map the nonholonomic Poisson bracket to the standard Poisson bracket on the cotangent bundle of the sphere. It means that any integrable system on the sphere has an integrable counterpart with respect to the nonholonomic bracket and vise versa. The list of the known integrable systems on the sphere can be found in \cite{bog86,bm05,rw85}.

In the next section we prove that we can not identify the Euler top and the nonholonomic Chaplygin ball using this mapping because they have different separated relations even at $(\g,M)=0$.

\subsection{Separation relations }
Substituting variables $\g,M$ (\ref{ftrans-ell}-\ref{ftrans-nell})   into the
St\"ackel potentials (\ref{st-pot}), we obtain a pair of separation  relations (\ref{stseprel}) for the Euler top and the  Chaplygin ball.  These separated equations define some algebraic curves and we can say that the equations of motion (\ref{m-eq}) are linearized on the symmetrized product of these curves.
\begin{proposition}
In holonomic case at $\omega=\mathbf AM$  the variables of separation lie on  two copies of the hyperelliptic curve
of genus one
\begin{equation}\label{s-relE}
\mathcal C^{(1)}:\qquad 4(a_1-\mathrm x)(a_2-\mathrm x)(a_3-\mathrm x)\,\mathrm y^2-(\mathrm x H_2-H_1)=0,\end{equation}
where $\mathrm x=u,v$ and $\mathrm y=p_u,p_v$.

In nonholonomic case at $\omega=\mathbf A_d M$ the variables of separation lie on  two copies of the following hyperelliptic curve of genus two
\begin{equation}\label{s-relC}
{\mathcal C}^{(3)}:\qquad 4(1-d\mathrm x)(a_1-\mathrm x)(a_2-\mathrm x)(a_3-\mathrm x)\,\mathrm y^2-(\mathrm x H_2-H_1)=0,\end{equation}
where $ \mathrm x=\mathrm{u},\mathrm{v}$ and $\mathrm y=\mathrm{p}_u,\mathrm{p}_v$.
\end{proposition}

Initial variables as functions on variables of separation  are given by (\ref{ftrans-ell}) and (\ref{ftrans-nell}).
It allows us to express  integrals of motion $H_{1,2}$ (\ref{int-h}) in terms of variables of separation. Substituting the resulting formulae for $H_{1,2}$ into the separated relation  we prove this proposition.
 $\square$

In fact, we obtain the variables of separation and the separated equations geometrically, i.e. without the equations of motion, the time variable and the underlying Hamiltonian or conformally Hamiltonian structures. We only suppose that the  foliation defined by the integrals $H_{1,2}$ (\ref{int-h}) on symplectic leaves of the corresponding Poisson brackets is  bi-lagrangian foliation.

However, in order to get the solutions of the separated equations $\mathrm x(t)$ and $\mathrm y(t)$
we have to explicitly introduce a time variable $t$. Solving separated equations with respect to $H_{1,2}$ one gets the Hamilton functions for the Euler top
\begin{equation}\label{ham-uv1}
H_1=\dfrac{4v(a_1-u)(a_2-u)(a_3-u)}{u-v}\,p_u^2+
\dfrac{4u(a_1-v)(a_2-v)(a_3-v)}{v-u}\,p_v^2,
\end{equation}
and for the Chaplygin ball
\begin{eqnarray}\label{ham-uv3}
H_1&=&\dfrac{4\mathrm{v}(1-d\mathrm{u})(a_1-\mathrm{u})(a_2-\mathrm{u})(a_2-\mathrm{u})}{\mathrm{u}-\mathrm{v}}\,\mathrm p_u^2\\
\nonumber\\
&+&\dfrac{4\mathrm{u}(1-d\mathrm{v})(a_1-\mathrm{v})(a_2-\mathrm{v})(a_3-\mathrm{v})}{\mathrm{v}-\mathrm{u}}
\,\mathrm p_{v}^2\,.
\nonumber
\end{eqnarray}
By definition the variables of separation are canonical variables and, therefore,
we have
\begin{equation}\label{pbr-uv} \begin{array}{ll}
\{H_1,\mathrm x_1\}=\dfrac{ 4\mathrm{x_2}\sqrt{\mathrm P(\mathrm x_1)}}{\mathrm{x}_1-\mathrm{x}_2}\qquad&
\{H_1,\mathrm x_2\}=\dfrac{ 4\mathrm{x_1}\sqrt{\mathrm P(\mathrm x_2)}}{\mathrm{x}_2-\mathrm{x}_1}\,,\\
\\
\{H_1,\mathrm x_1\}_d=\dfrac{ 4\mathrm{x_2}\sqrt{\mathrm P_d(\mathrm x_1)}}{\mathrm{x}_1-\mathrm{x}_2}\qquad&
\{H_1,\mathrm x_2\}_d=\dfrac{ 4\mathrm{x_1}\sqrt{\mathrm P_d(\mathrm x_2)}}{\mathrm{x}_2-\mathrm{x}_1}\,.
\end{array}
\end{equation}
Here variables $\mathrm x_{1,2}$ are coordinates of separation  $u,v$ or $\mathrm u, \mathrm v$, respectively. Polynomials  $\mathrm P(\mathrm x)$ and $\mathrm P(\mathrm x)$ are the polynomials of degree $4$ and $5$ in $\mathrm x$ variable
\begin{eqnarray}
\mathrm P(\mathrm x)&=&(a_1-\mathrm x)(a_2-\mathrm x)(a_3-\mathrm x)(\mathrm x H_2-H_1),\nonumber\\
\nonumber\\
\mathrm P_d (\mathrm x)&=&(1-d\mathrm x)(a_1-\mathrm x)(a_2-\mathrm x)(a_3-\mathrm x)(\mathrm x H_2-H_1).\nonumber
\end{eqnarray}
 On the other hand, according to (\ref{new-eq}), the  brackets (\ref{pbr-uv}) are equal to
\[\{H_1,\mathrm x_{1,2}\}=\dfrac{2}{\mu}\,\dfrac{d\mathrm x_{1,2}}{dt}\qquad\mbox{and}\qquad \{H_1,\mathrm x_{1,2}\}_d=\dfrac{2}{\mu_d}\,\dfrac{d\mathrm x_{1,2}}{dt}\]
where
\[
\mu=1\,,\qquad\mbox{and}\qquad \mu_d=\sqrt{\mathrm g(\g)}=\sqrt{\dfrac{(1-d\mathrm u)(1-d\mathrm v)}{(1-da_1)(1-da_2)(1-da_3)}}
\]
are Jacobi multipliers (\ref{jac-mul1}-\ref{jac-mul3}).

For the Chaplygin ball, in order to get the solutions $\mathrm x_{1,2}(t)$ of the equations of motion,
we have to consider the Jacobi inversion problem for the equations
\begin{eqnarray}
\beta_1-2\int \mu_d\,dt&=& \int \frac{d \mathrm x_1}{\sqrt{\mathrm P_d (\mathrm x_1)}}
  +\int \frac{d \mathrm x_2}{\sqrt{\mathrm P_d (\mathrm x_2)}},\nonumber\\
\label{quadr}\\
\beta_2&=& \int \frac{\mathrm x_1 d \mathrm x_1}{\sqrt{\mathrm P_d (\mathrm x_1)}}
  +\int \frac{\mathrm x_2 d \mathrm x_2}{\sqrt{\mathrm P_d (\mathrm x_2)}},\nonumber
\end{eqnarray}
where  $\beta_{1,2}$ are the constants of integration. The change of time variable (\ref{t-ch}) reduces these equations to the standard Abel-Jacobi equations \cite{ch03,koz85}.

It is easy to prove that the right hand side in  $\beta_2$ (\ref{quadr}) coincides with an additional Euler-Jacobi quadrature emerged in the Jacobi last multiplier theory. Of course, for the Chaplygin ball this quadrature can be obtained without any change of time variable.

\section{Generalizations of the nonholonomic Chaplygin ball at  $C_2=0$.}

Equations of motion, Poisson brackets, Lax matrices and  classical $r$-matrices for the  Chaplygin ball are  deformation of the same objects for the Euler top by parameter $d$. Moreover, at $C_1=1$ and $C_2=0$  we know how to deform the corresponding variables of separation and the separated relations.   Because  at $C_1=1$ and $C_2=0$ our phase space  is equivalent to  the cotangent bundle $T^*\mathbb S$ of the unit two-dimensional Poisson sphere, we can obtain similar deformations of other  integrable systems on cotangent bundles of the Riemannian manifolds.

\subsection{The $2\times 2$ Lax matrices.}
In variables of separation we deal with the uniform St\"ackel systems (\ref{ham-uv1}-\ref{ham-uv3}) and, therefore,  we can get $2\times 2$ Lax matrices associated with the Abel-Jacobi equations (\ref{quadr}) in a standard  way, see \cite{eekt,kuz92,ts99,ts99a,ts04a} as well as the relevant references therein.

According to \cite{ts99,ts99a,ts04a}, let us introduce the following functions on  the canonical variables of separation and spectral parameter $\lambda$
\[h(\lambda)=-\dfrac{1}{8}\,\Bigl\{\,H_2,e(\lambda)\,\Bigr\}\,\qquad
h_d(\lambda)=-\dfrac{1}{8(1-d\lambda)}\,\Bigl\{\,H_2,e_d(\lambda)\,\Bigr\}_d\,,
\]
and
\begin{eqnarray}
f(\lambda)&=&\dfrac{1}{4}\left(\,\Bigl\{\,H_2,h(\lambda)\,\Bigr\} -e(\lambda)H_2\right)\,,\nonumber\\
\nonumber\\
f_d(\lambda)&=&\dfrac{1}{4(1-d\lambda)}\left(\,\Bigl\{\,H_2,h_d(\lambda)\,\Bigr\}_d
-\left(1+\dfrac{\mbox{\rm tr} \mathbf A-2(\mathrm u+\mathrm v)}{1-2d\lambda}\right)e_d(\lambda)H_2\right.\nonumber\\
\nonumber\\
&&\qquad\qquad\qquad\qquad\qquad\qquad-\left.\dfrac{e_d(\lambda)H_1}{1-d\lambda}
\right)\nonumber\,.
\end{eqnarray}
Here $e(\lambda)$  and $e_d(\lambda)$ are given by (\ref{ell-q}-\ref{nhell-q}).

For the brevity below we will use the following  denotation, index $x$ is a white space for the Euler top and its Hamiltonian generalizations, whereas $x=d$ for the Chaplygin ball and the corresponding generalizations.
\begin{proposition}
 At $(\g,M)=0$  the Lax matrices
\begin{equation}\label{lax2}
\mathscr{L}_x =\left(
             \begin{array}{cc}
               h_x  & e_x  \\
               f_x  & -h_x
             \end{array}
           \right),\qquad
\mathscr{A}_x =\dfrac{1}{\mu_x \,e_x }\left(
                                       \begin{array}{cc}
                                         -{e}_x ^\prime & 0 \\
                                        2 {h}_x ^\prime & {e}_x ^\prime
                                       \end{array}
                                     \right),
\end{equation}
satisfy to the Lax equation
\begin{equation}\label{laxeq-22}
\dfrac{d}{dt}\,\mathscr L_x (\lambda)=\dfrac{\mu_x }{2}\,\Bigl\{H_1,\mathscr L_x \Bigr\}_x =\Bigl[\mathscr L_x (\lambda),\mathscr A_x (\lambda)\Bigr]\,.
\end{equation}
Here ${z}^\prime=\{z,H_1\}_{x}$ is a time derivative up to Jacobi multiplier  (\ref{jac-mul1}-\ref{jac-mul3}).
\end{proposition}
 The proof is a straightforward verification of the equations  (\ref{laxeq-22}) for the given Lax matrices.
 $\square$

As usual, substituting $\lambda=\mathrm x$ into the determinants of the  Lax matrices
\[\det \mathscr L_x (\lambda)=-h_x^2(\lambda)-e_x(\lambda)f_x(\lambda)\,,\]
which are equal to
\begin{eqnarray}
\det\mathscr L(\lambda)&=&-\dfrac{\lambda H_2-H_4}{4(a_1-\lambda)(a_2-\lambda)(a_1-\lambda)}\,,\nonumber\\
\nonumber\\
\det\mathscr L_d (\lambda)&=&-\dfrac{\lambda H_2-H_4}{4(1-d\lambda)(a_1-\lambda)(a_2-\lambda)(a_1-\lambda)}\,,\nonumber\
\end{eqnarray}
one gets separated relations (\ref{s-relE}) and (\ref{s-relC}) because  $e_x(\mathrm x)=0$ and $h_x(\mathrm x)=\mathrm y$.

In \cite{ch03} Chaplygin reduces the generic case  at $(\g,M)\neq0$ to the particular case  at $(\g,M)=0$. By applying the inverse map to the Lax matrices (\ref{lax-ch}) one gets the Lax matrices for the generic case of the nonholonomic Chaplygin ball. These matrices and the corresponding $r$-matrix algebra will be studied in a forthcoming separate publication.

Matrices $\mathscr L_k(\lambda)$ are associated with the uniform St\"{a}ckel systems and, therefore, they satisfy to the linear $r$-matrix algebra (\ref{rrpoi}) with the well-studied dynamical $r$-matrices \cite{eekt,kuz92,ts99,ts99a,ts04a}.
In contrast with the previous $3\times 3$ Lax matrices (\ref{lax-ch}) it allows us to obtain some well studied generalizations of these $2\times 2$ matrices in the next paragraphs.

\subsection{Chaplygin ball and separable potentials.}

We are going to demonstrate that the Chaplygin ball at $(\g,M)=0$  is still integrable in the force fields associated with a huge family of the so-called separable potentials \cite{bog86,eekt,rw85}.

It is well known of how to get various generalizations of the separable systems
using the deformations of their separated equations \cite{jac66}. For instance, let us consider the following  deformations of the separation relations  (\ref{s-relE}) and (\ref{s-relC})
\[
4(a_1-\mathrm x)(a_2-\mathrm x)(a_3-\mathrm x)\,\mathrm y^2-(\mathrm x H_2-H_1)+V(\mathrm x)=0
\]
or
\[
4(1-d\mathrm x)(a_1-\mathrm x)(a_2-\mathrm x)(a_3-\mathrm x)\,\mathrm y^2-(\mathrm x H_2-H_1)+V(\mathrm x)=0\,,
\]
where the potential $V$ is some function on $\mathrm x$. Usually, potential $V$ is a linear combination of the trivial separable potentials  $V_m=\alpha_m\mathrm x^m$, where $m$ is a positive or negative integer \cite{bog86,rw85}.

In order to get the same deformations in the initial variables $\g,M$ we can use the generating function \cite{rw85}
\[
\Phi(\lambda)=\dfrac{\phi(\lambda)}{e_x (\lambda)}\,,
\]
or the determinant of the corresponding deformations the Lax matrix $\mathscr L_x (\lambda)$ (\ref{lax2})
\[
f_x \to f_x +\left[\dfrac{\phi(\lambda)}{e_x (\lambda)}\right]_{MN}\,.
\]
Here $\phi(\lambda)$ is a parametric function on spectral parameter and
$[\xi(\lambda)]_{MN}$ is a linear combination of the Laurent projections of $\xi(\lambda)$ by $\lambda$ \cite{eekt,ts99,ts99a}.

For example, if $V=\alpha\mathrm x^2$ one gets the integrable system
\begin{equation}\label{int-neum}
H_2=(M,M)+{\alpha}(\g,\mathbf A\g)\,,\qquad
H_1=(M,\mathbf A_1M)-\dfrac{\alpha}{a_1a_2a_3}(\g,\mathbf A^{-1}\g)\,,
\end{equation}
which can be identified with the Neumann system on the sphere, and its nonholonomic counterpart
\begin{eqnarray}
H_2&=&(M,M)+\alpha \mathrm g(\g)\,\Bigl((\g,\mathbf A \g)-d(\mathbf A \g,\mathbf A\g)\Bigr)\,,\label{int-nneum}\\
H_1&=&(M,\mathbf A_3M)-\alpha\,\mathrm g(\g)\,\,\left(
\dfrac{(\g,\mathbf A^{-1}\g)}{a_1a_2a_3}-da_1a_2a_3\right)\,.\nonumber
\end{eqnarray}

Another nonholonomic analog of the Neumann system with the polynomial in $\g$ potential has been proposed by Kozlov  \cite{koz85}
at
\begin{equation}\label{pol-neum}
V=-\alpha\mathrm x^2+(a_1+a_2+a_3)\mathrm x+\dfrac{\alpha(a_1-\mathrm x)(a_2-\mathrm x)(a_3-\mathrm x)}{1-d\mathrm x}\,,
\end{equation}
see integrals of motion in (\ref{int-cl}). At $(\g,M)=0$ this system is separable in the Chaplygin coordinates \cite{fed85}.

If $V=\beta\mathrm x^3$ we obtain a forth order polynomial potential on the sphere
\begin{equation}\label{4order-spt}
H_1=(M,\mathbf A_1M)+\beta\dfrac{(\g,\mathbf A^{-1}\g)}{a_1a_2a_3}\Bigl( (\g,\mathbf A\g)-\mbox{tr}\mathbf A  \Bigr)\,,
\end{equation}
  and its nonholonomic analog
\begin{eqnarray}\label{4order-nhspt}
H_1&=&(M,\mathbf A_3M)+{\beta\,\mathrm g(\g)}\left(\dfrac{(\g,\mathbf A^{-1}\g)}{a_1a_2a_3}-d{a_1a_2a_3}\right)\times\nonumber
\nonumber\\
&\times&\left[
\mathrm g(\g)\,\Bigl((\g,\mathbf A \g)-d(\mathbf A \g,\mathbf A\g)\Bigr)-\mbox{tr}\mathbf A\right]
\,.
\end{eqnarray}
Similarly  we can  get other well-known integrable systems on the sphere \cite{bog86,rw85}, such as Braden  and Rosochatius systems, and their nonholonomic counterparts separable in the nonholonomic elliptic coordinates.

\subsection{Deformations of natural Hamiltonian systems on Riemannian spaces of constant curvature.}

At $(\g,M)=0$ the Euler top is  a  dynamical system describing free motion on the two-dimensional  sphere, which may be identified  with a  particular case of  the Gaudin magnet \cite{kuz92}. It is well-known how to describe similar $N$-dimensional integrable systems on any Riemannian space  of constant curvature   and  then how to add separable potentials to these systems, see  \cite{eekt,kal,kuz92,ts99,ts99a,ts04,ts04a} and references within. There are many different tools to investigations of such systems \cite{bog86,bm05,rs}. Below we will use only one of them based on separation of variables method.

The key ingredient of this construction is $2\times 2$ Lax matrix associated with the $sl(2)$ Gaudin magnet  \cite{eekt,kuz92,ts04,ts04a}, which  is completely defined by the rational function $e(\lambda)$ on a given Riemannian manifold. The list of all admissible functions may be found in \cite{kal}.

Let us start with elliptic coordinates $(q,p)$ on the sphere $\mathbb S_{N}$ in $N+1$-dimensional Euclidean space $\mathbb R_N$. In this case deformation of the free motion similar to Chaplygin ball consists of three steps:
\begin{itemize}

\item we have to  change the  separation relations from
\begin{equation}\label{s-relEm}
 \prod_{j=1}^{N+1}(a_j-q_i)\,p_i^2-(q_i^{N-1}H_N+\cdots+q_iH_2+H_1)=0\,, \quad i=1,\ldots,N,
\end{equation}
to
 \begin{equation}\label{s-relCm}
 (1-dq_i)\prod_{j=1}^{N+1}(a_j-q_i)\,p_i^2-(q_i^{N-1}H_N+\cdots+q_iH_2+H_1)=0\,, \qquad\qquad
\end{equation}
compare with (\ref{s-relE}-\ref{s-relC});
\item we have to change the definition of $q_j$ in term of cartesian coordinates in $\mathbb R_N$
from \begin{equation}\label{ell-sph}
e(\lambda)=\sum_{k=1}^{N+1}\dfrac{{\g}_k^2}{\lambda-a_k}=
\dfrac{\prod_{i=1}^{N}(\lambda-q_i)}{\prod_{j=1}^{N+1}(\lambda-a_j)}\,,
\end{equation}
that implies $\sum_{i=1}^{N+1} {\g}_i^2=1$, to
\begin{equation}\label{ell-sphC}
e_d(\lambda)=\mathrm g(\g) \sum_{k=1}^{N+1}\dfrac{\g_k^2\,(1-da_k)}{\lambda-a_k}=
\dfrac{\prod_{i=1}^{N}(\lambda-q_i)}{\prod_{j=1}^{N+1}(\lambda-a_j)}\,,
\end{equation}
where $\mathrm g(\g)$ is defined by residue of $e_d(\lambda)$ at infinity;
\item we have to change the time variable in order to attach some nonholonomic physical meaning to the proposed pure mathematical integrals $H_1,\ldots,H_N$ and the corresponding equations of motion, see \cite{jov09,ho09}.
\end{itemize}
Remind that  rewriting  separated relations (\ref{s-relEm}) or  (\ref{s-relCm})  in the St\"ackel form (\ref{stseprel})
\[
\Phi_i=p_i^2+\sum_{j=1}^m S_{ij}(q_i)H_j=0,\qquad i=1,\ldots,m,
\]
we can easily determine  integrals of motion in a standard way \cite{ts99,ts99a}
\[
H_k=\sum_{i=1}^m C_{ki}\,p_i^2\,,
\]
where $C=S^{-1}$ is the inverse matrix to the St\"{a}ckel matrix $S$, which is  a standard  transpose Brill-Noether matrix with entries divided by  $ \prod_{j=1}^{N+1}(a_j-q_i)$ or $ (1-dq_i)\prod_{j=1}^{N+1}(a_j-q_i)$. The same integrals may be rewritten in the following form  \cite{ts04a}
\[\left.H_k=\mbox{res}\right|_{\lambda=\infty} \,\lambda^{N-k}e^{-1}(\lambda)\,.\]
Of course, instead of  one parametric deformation  we can  consider multi-parameter deformations replacing terms $(1-dq_j)$ and $(1-da_j)$ on  $\prod(1-d_mq_j)$   and  $\prod(1-d_ma_k)$  in (\ref{s-relCm}) and (\ref{ell-sphC}), respectively.

On the other hand instead of (\ref{ell-sph}) we can start with any other coordinate system and the corresponding separated equations on the Riemannian spaces of constant curvature \cite{kuz92,ts04a}. For instance, we can take elliptic  coordinates
\begin{equation}\label{ell-c}
e(\lambda)=1+\sum_{k=1}^N\dfrac{\g_k^2}{\lambda-e_k}=
\dfrac{\prod_{j=1}^N(\lambda-q_j)}{\prod_{i=1}^N(\lambda-e_i)}\,.
\end{equation}
or parabolic coordinates
\begin{equation}
e(\lambda)=\lambda-2\g_N-\sum_{k=1}^{N-1}\dfrac{\g_k^2}{\lambda-e_k}=
\dfrac{\prod_{j=1}^N(\lambda-q_j)}{\prod_{i=1}^{N-1}(\lambda-e_i)}
\end{equation}
in  $N$-dimensional Euclidean space.

In order to consider systems with potential we can add  St\"ackel potential $U(q_i)$ to  $i$-th separated equation (\ref{s-relEm}) and (\ref{s-relCm}).

For all these deformations we easy calculate  $2\times 2$ Lax matrices in terms of variables of separation, because all these deformations are uniform St\"ackel systems associated with various hyperelliptic curves  \cite{kuz92,ts99,ts99a, ts04,ts04a}

So, there are not mathematical problems in the  construction of such ``nonholonomic" dynamical systems associated with any  orthogonal coordinate system and their potential generalizations. The main problem is definition of a suitable change of time variable, which may be  associated with an interesting  physical model.

\section{Generalizations of the nonholonomic Chaplygin ball at $(\g,M)\neq0$.}
In the previous section we consider various deformations of our dynamical systems at  $(\g,M)=0$ using  the variables of separation method for the Hamiltonian systems  and the suitable time reparametrization. We proceed by discussing some possible deformations of the Chaplygin ball in generic case.

As above index $x$ is a white space for generalizations of the Euler top and $x=d$ for generalizations of the Chaplygin ball.

Let us consider the deformation of the  equations (\ref{m-eq})
\begin{equation}\label{meq-v}
\dot M=M\times \omega+\g\times b\,,
\qquad
\dot \g=\g\times \omega\,,
\end{equation}
where $\omega=\mathbf A_{x}M$ and vector $b$ is an arbitrary  function on $\g$ and $M$. We want to discuss  only an existence of integrals of motion in involution with respect to the Poisson bracket  $\{.,.\}_d$, which is a necessary condition for the Liouville integrability of the corresponding Hamiltonian systems. The  invariant measures and the corresponding  time transformations are considered in review \cite{bm08}.

It is clear that $\mathcal H_{1,2}$ (\ref{3int}) remain the integrals of equations (\ref{meq-v}) and upon the same basis  we can identify these integrals with Casimir functions (\ref{caz}) on our Poisson manifold. It allows us to look for two additional integrals $\mathcal H_3$ and $\mathcal H_4$ in  involution with respect to the same Poisson brackets $\{.,.\}$ (\ref{1-br}) and
 $\{.,.\}_d$ (\ref{3-br})
 \[
 \{\mathcal H_3,\mathcal H_4\}=0\,,\qquad  \{\mathcal H_3,\mathcal H_4\}_d=0\,,
 \]
 where $\{.,.\}_d$ is  the ``nonholonomic'' deformation of canonical bracket $\{.,.\}$ on $e^*(3)$.  For the first Poisson bracket all possible integrable deformations are well known \cite{au96,bm05,rs}.  So, we can try to get ``nonholonomic'' deformations of the  Lagrange and Kowalevski tops,  or of the Kirchhoff, Clebsh and Steklov-Lyapunov systems.

If the Hamilton function reads as
\[
2H=\mathcal H_4=(M,\omega)+2V(\g)\,,\qquad \omega=\mathbf A_x M,
\]
then in holonomic case equations (\ref{meq-v}) are identified with the Euler-Poisson equations \cite{au96,bm05}
\begin{equation}\label{ham-eq1}
\dot M=M\times \dfrac{\partial H}{\partial M}+\g\times \dfrac{\partial H}{\partial \g}\,,\qquad
\dot \g=\g\times \dfrac{\partial H}{\partial M}\,,\qquad H=\dfrac12\, \mathcal H_4,
\end{equation}
whereas in nonholonomic case first equation has to be replaced to
\begin{equation}\label{ham-eq2}
\dot M=M\times \dfrac{\partial H}{\partial M}+\g\times \dfrac{\partial V}{\partial \g}\,,
\end{equation}
according to the procedure of elimination of the undetermined
Lagrange multipliers, see review \cite{bm08} and references within.

\subsection{Linear integrals of motion.}

Let us briefly consider the Lagrange top \cite{au96,bm05,rs} and its nonholonomic twin \cite{ch97,gal04}.
Recall that Lagrange top  is a special case of rotation of a rigid body around a
fixed point in a homogeneous gravitational field, characterized by the following conditions:
the rigid body is rotationally symmetric, i.e. two of its three principal moments of
inertia coincide, and the fixed point lies on the axis of rotational symmetry.
In much the same way second system is the Chaplygin ball with the rotationally symmetric mass distribution
in the  homogeneous gravitational field.

\begin{proposition}
If $\omega=\mathbf A_x M$,
\[
a_{1}=a_{2},\qquad\mbox{and}\qquad b=(0,0,b_3)\,,
\]
then the integrals of the equations (\ref{ham-eq1})-(\ref{ham-eq2})
\begin{equation}\label{int-lag}
\mathcal H_3=(M,M)+2a_1^{-1}(b,\g)\qquad\mbox{and}\qquad \mathcal H_4=(M,\omega)+2(b,\g),\qquad
\end{equation}
are in the involution with respect to the Poisson brackets $\{.,.\}$ (\ref{1-br}) and $\{.,.\}_d$ (\ref{3-br}), respectively.
\end{proposition}
The proof is a straightforward calculation of the Poisson brackets between integrals of motion.

In holonomic case the linear in momenta integral
\[K=(b,M)=M_3,\qquad \{\,\mathcal H_{k},K\,\}=0,\qquad k=1,2,3,\]
 can be obtained from the quadratic integrals (\ref{int-lag}) in a standard way
\begin{equation}\label{lin-lag}
\sqrt{\mathcal H_4-a_1\mathcal H_3}=\sqrt{a_3-a_1}M_3=\sqrt{a_3-a_1}K.
\end{equation}
In nonholonomic case the linear integral  looks like
\begin{equation}
K=\sqrt{\mathrm g(\g)}\,\left(M_3+\dfrac{a_1x_3(\g,M)}{1-da_1(\g,\g)}\right)\,,\qquad
\{\,\mathcal H_{k},K\,\}_d=0\,\quad k=1,2,3\,.
\end{equation}
It can be represented via quadratic integrals (\ref{int-lag}) and the Casimir functions according to the relation
\[
\mathcal H_4-a_1\mathcal H_3+(a_1-a_3)\bigl(1-da_1(\g,\g)\bigr)\,K^2-\dfrac{a_1^2}{1-da_1(\g,\g)}\, (\g,M)^2=0\,.
\]
In  both cases the equations of motion (\ref{meq-v}) are equivalent to the  Lax  equation
\[
\dfrac{d\mathbf L}{dt}=[\mathbf L,\mathbf \Omega+\lambda \mathbf B],\qquad \mathbf L=\mathbf M+\dfrac{\mathbf \Gamma}{\lambda},\qquad\qquad \lambda\in\mathbb R\,.
\]
It is obvious, that the Lax matrix $\mathbf L$ satisfies the linear $r$-matrix brackets (\ref{rrpoi}) with the same $r$-matrices (\ref{r-eul}) and (\ref{r-ch}).

In holonomic case by $\omega=\mathbf A_1M$  we have $b\times (M-\omega)=0$ at $a_1=a_2=1$ and $b_1=b_2=0$. It allows us to get another well-known Lax  representation for the Lagrange top \cite{au96,rs}:
\begin{equation}\label{lax-lag}
\dfrac{d\mathbf L}{dt}=[\mathbf L,\mathbf \Omega+\lambda \mathbf B],\qquad \mathbf L=\lambda \mathbf B+\mathbf M+\dfrac{\mathbf \Gamma}{\lambda}\,.
\end{equation}
In nonholonomic case $b\times (M-\omega)\neq 0$ and we have no such Lax matrix at all. Of course, it is a superficial argument because the main point is that the nonholonomic system is related with the genus three algebraic curve instead of the elliptic curve for the Lagrange top.

Namely, using the Euler angles and their conjugated momenta, for the Lagrange top we can easily prove  that  the pair of  canonical variables
\[u = \g_3=\cos(\theta),\qquad p_u = \dfrac{\g_2 M_1-\g_1M_2}{\g_1^2+\g_2^2}=-\dfrac{p_\theta}{\sin(\theta)},\qquad \{u,p_u\}=1,
\]
lies on the elliptic curve  defined by equation
\[
a_1p_u^2+\dfrac{2ub+\beta-a_3\, \alpha^2}{u^2-1}+\dfrac{a_1(\alpha u+\ell)^2}{(u^2-1)^2}
=0\,,
\]
where we fix the values of the  integrals of motion
\begin{equation}\label{int-vallag}
\mathcal H_1=(\g,\g)=1,\qquad \mathcal H_3=(\g,M)=\ell,\qquad K=\alpha,\qquad \mathcal H_4=\beta\,.
\end{equation}

For the nonholonomic system canonical variables
\[
 u=u,\qquad \hat{p}_u=\dfrac{p_u}{\sqrt{\mathrm g}},\qquad \{u,\hat{p}_u\}_d=1,
\]
satisfy to the following separated equation
\begin{equation}\label{lag-nh}
a_1\hat{p}_u^2+\dfrac{2ub+\beta-a_3 \rho\,\alpha^2}{u^2-1}\,\mathrm g+\dfrac{a_1\,(\alpha \,u+\varrho\,\ell)^2}{\bigl(1+da_1(u^2-1)\bigr)(u^2-1)^2}=0\,.
\end{equation}
Here we fix the values of the integrals of motion as in (\ref{int-vallag}) and
\begin{eqnarray}
\mathrm g&=&\dfrac{d}{1-da_1+d(a_1-a_3)u^2}
\,,\qquad
\rho=\dfrac{(1-da_1)^2}{1+da_1(u^2-1)}\,,
\nonumber\\
\nonumber\\
\varrho&=&-\dfrac{1+da_1(u^2-1) }{1-da_1}\sqrt{\mathrm g}\,.\nonumber
\end{eqnarray}
If $\ell=(\g,M)=0$,  one gets the elliptic curve, but in generic case
rewriting  the equation (\ref{lag-nh}) in polynomial form we obtain the algebraic
curve of genus three.

Only at $(\g,M)=0$  we can get the solutions in the terms of elliptic functions and, therefore, only in this particular case we can try to reconstruct the Lax matrix associated with the elliptic curve.

Three different bi-Hamiltonian structures for the  Lagrange top have been obtained in \cite{ts08l}. These structures are related with different variables of separation and, therefore, different quadratures. If we get similar dynamical Poisson bivectors for its nonholonomic counterpart, one gets various quadratures, which could be associated with the distinct Lax matrices  and underlying $r$-matrix algebras.

\subsection{Second order integrals of motion}

For the Kirchhoff problem, the integrable cases by Kirchhoff, Clebsch, and Steklov-Lyapunov
are known. In this section we begin with the Clebsch case
\begin{proposition}
If $\omega=\mathbf A_x M$ then  the integrals of the  equations (\ref{ham-eq1})-(\ref{ham-eq2})
\begin{equation}\label{int-cl}
\mathcal H_3=(M,M)-(\mathbf A\g,\g)\qquad\mbox{and}\qquad \mathcal H_4=(M,\omega)+(\mathbf A^\vee\g,\g)
\end{equation}
are in the involution with respect to the  Poisson  brackets $\{.,.\}$ (\ref{1-br}) and $\{.,.\}_d$ (\ref{3-br}), respectively.
\end{proposition}
As above proof is a straightforward calculation of the Poisson brackets between integrals of motion.

In holonomic case we have the well-studied  Clebsch problem. In nonholonomic case this deformation of the Chaplygin system has been proposed by Kozlov  \cite{koz85} in framework of the Euler-Jacobi integration procedure, i.e. without notion of  the  Poisson bracket.

There are some different Lax matrices for the Clebsch model \cite{bm05,rs}. For example,
\[
\mathbf L=\lambda \mathbf A+\mathbf M+\dfrac{\g\times \g}{\lambda}\,.
\]
We can not directly generalize this matrix to the nonholonomic case, because  $\dot{\mathbf A}_d\neq 0$ (\ref{dot-a}) in contrast with $\dot{\mathbf A}=0$ above.  We suppose that the nonholonomic Kozlov system is related with the algebraic curve of higher genus and, therefore, the corresponding Lax matrices will be more complicated deformations of the known Lax matrices for the Clebsch problem.

At $(\g,M)=0$ the Clebsch system becomes the so-called Neumann system on the sphere, which is separable in the elliptic coordinates $u,v$ (\ref{ell-q}) \cite{bm05}. Its nonholonomic counterpart is the separable system in Chaplygin coordinates $\mathrm u,\mathrm v$ (\ref{nhell-q}) \cite{fed85} and, therefore, we can get $2\times 2$ Lax matrices $\mathscr L(\lambda)$ (\ref{lax2}) for this nonholonomic system as well.

The Clebsch case is equivalent to the Brun case of integrability in the Euler-Poisson equations \cite{bm05} and, moreover,  it is  trajectory isomorphic to the Kowalevski gyrostat \cite{kom05}. We can hope to get a nonholonomic analog of the Kowalevski top by using similar isomorphism.

Now let us briefly discuss  the integrable Steklov-Lyapunov case in the Kirchhoff equation and
the corresponding integrals of motion
\begin{equation}\label{int-sl}
\begin{array}{l}
\mathcal H_3=(M,M)-2(M,\mathbf A\g)+(\g,\mathbf C^2 \g)\,,\\
\\
\mathcal H_4=(M,\omega)+2(M,\mathbf A^\vee\g)+(\mathbf A\g,\mathbf C^2\g)\,,\end{array}
\end{equation}
where  $\mathbf C=\mbox{\rm diag}(a_2-a_3,a_3-a_1,a_1-a_2)$.
These integrals
are in the involution with respect to the first bracket $\{\mathcal H_3,\mathcal H_4\}=0$.

If we replace $\omega=\mathbf AM$ on $\omega=\mathbf A_d M$ in $\mathcal H_4$
then $\{\mathcal H_3,\mathcal H_4\}_d =0$ only if two parameters $a_i$ coincide with each other. So, for the nonholonomic bracket $\{.,.\}_d$ we have to propose some more complicated deformations of the integrals of motion (\ref{int-sl}).

It is known that the Steklov-Lyapunov system is equivalent to the integrable system on the sphere with forth order potential (\ref{4order-spt}) \cite{ts04}. We suppose that a similar transformation of the system
(\ref{4order-nhspt}) separable in nonholonomic elliptic coordinates allows us to get a nonholonomic counterpart of the Steklov-Lyapunov system.

\section{Conclusion}

We consider two very similar dynamical systems, which evolve on coadjoint orbits of Lie algebra $e(3)$ and their non-trivial symplectic deformations.

Close ties between the integrable Euler top and the nonholonomic Chaplygin ball allow us to get Lax matrices, $r$-matrices and bi-hamiltonian structure for this nonholonomic system. Moreover, in framework of the Jacobi method of separation of variables we describe a huge
family of separable potentials, which can be added to nonholonomic Hamiltonian and briefly discuss how to get the $N$-dimensional nonholonomic systems on the Riemannian spaces of constant curvature.

In \cite{ch03} Chaplygin transforms the generic case of the rolling ball to the particular case  of horizontal angular momentum $(\g,M)=0$.  It allows us to solve the equations of motion using the same variables of separation $\mathrm u,\mathrm v$ (\ref{nhell-q}), which  will be the non-canonical variables with respect to initial Poisson bracket $\{.,.\}_d$ (\ref{3-br}) after this map.  We will  discuss this Chaplygin map in framework of the Poisson geometry in separate publication, as well as the corresponding $2\times 2$ Lax matrices and the underlying $r$-matrix algebra.

\section*{Acknowledgments}
We would like to thank  A.V. Borisov, Yu.N. Fedorov, I.S. Mamaev and A.A. Kilin  for gentle introduction to the nonholonomic theory and referees for useful suggestions and pointing some references.

\end{document}